\documentclass[twocolumn,prl,superscriptaddress,showpacs]{revtex4}
\usepackage{amsmath,amssymb,graphicx}

\newcommand{\olcite}[1]{Ref.~\onlinecite{#1}}

\newcommand{\etat}{ \eta_{\rm r}^{\rm per}}
\newcommand{\rhot}{ \rho^{\rm per}}
\newcommand{\etas}{ \eta_{\rm s}}
\newcommand{\etar}{ \eta_{\rm r}}
\newcommand{\nt}{ n^{\rm per}}
\newcommand{\vex}{v_{\rm ex}}
\newcommand{\vh}{v_{\rm h}}
\newcommand{\Vc}{V_{\rm c}}
\newcommand{\pid}{p_{\rm id}}

\begin{document}

\title{Depletion-induced percolation in networks of nanorods}

\author{T.~Schilling}
\affiliation{Institut f\"ur Physik, Johannes Gutenberg-Universit\"at,
D-55099 Mainz, Staudinger Weg 7, Germany}

\author{S.~Jungblut}
\affiliation{Institut f\"ur Physik, Johannes Gutenberg-Universit\"at,
D-55099 Mainz, Staudinger Weg 7, Germany}

\author{Mark A.~Miller}
\affiliation{University Chemical Laboratory, Lensfield Road,
Cambridge CB2 1EW, United Kingdom}

\pacs{81.07.De, 82.70.Dd}

\date{\today}

\begin{abstract}
Above a certain density threshold, suspensions of rod-like colloidal
particles form system-spanning networks. Using Monte Carlo simulations,
we investigate how the depletion forces caused by spherical particles affect
these networks in isotropic suspensions of
rods. Although the depletion forces are strongly anisotropic and 
favor alignment of the rods, the percolation threshold of the rods decreases 
significantly. The relative size of the effect increases with
the aspect ratio of the rods. The structural changes induced in the
suspension by the depletant are characterized in detail and the system
is compared to an ideal fluid of freely interpenetrable rods.
\end{abstract}

\maketitle

Networks of electrically conducting carbon nanotubes
(CNTs) have important applications in the development of
light-weight conducting composites and films
\cite{coleman:1998,sandler:1999,benoit:2002,potschke:2003,vigolo:2000}.
At equilibrium,
colloidal suspensions of CNTs contain clusters with a distribution
of sizes.  The contacts within each cluster extend conductivity beyond
the length of individual CNTs, and contiguous conducting paths first
appear on a macroscopic scale when the CNT density crosses the percolation
threshold, at which the average cluster size diverges.
The conductivity of the sample rises sharply by several
orders of magnitude at this point \cite{coleman:1998,sandler:1999,benoit:2002,
potschke:2003,vigolo:2000}. The large aspect ratio
of CNTs means that, for a disordered distribution of positions and
orientations, a single particle provides connectivity over a
longer distance than several more spherical particles with the same total
volume.  Hence, an insulating matrix can be loaded with a smaller volume
fraction of conducting filler, leading to a lighter material.
\par
The percolation threshold of the CNTs depends both on their aspect
ratio, and on the interactions between them.  Recent studies
show that the addition of surfactant to a suspension of CNT bundles
(CNTBs) substantially lowers the percolation threshold \cite{vigolo:2005}.
At sufficiently high concentrations, the surfactant self-assembles into
micelles, which have been shown to induce depletion attraction between
the bundles \cite{wang:2004}.  Due to the weakness
of the attraction, the percolation is reversible and the system remains
at thermodynamic equilibrium, in contrast to strongly interacting particles,
which may form structurally arrested gels \cite{mohraz:2004}.
\par
The depletion forces arise from competition between the translational and
orientational entropy of the CNTBs on the one hand, and the 
translational entropy of the micelles on the other.  If two bundles approach
each other closely, the overall volume available to the micelles increases,
thereby inducing an effective attraction. For rod-shaped particles
like CNTBs, this depletion effect is highly anisotropic, since the
total volume excluded to the depletant depends on the relative orientations
of the bundles \cite{li:2005, savenko:2006}.  
The excluded volume is minimized by parallel arrangements of
neighboring rods, but alignment also reduces the spatial extent
of a cluster, and should therefore suppress rather than
enhance the formation of percolating networks \cite{munson-mcgee:1991}.  
Even in systems of spheres, it is possible for attraction to raise the
percolation threshold \cite{bug:1985}.
Nevertheless, recent analytic work on mixtures of rods and spherical 
depletants predicts that depletion does lower the percolation
threshold of the rods for all depletant concentrations \cite{wang:2003}.
In the present study, we will see that depletion induces relatively local
changes in the structure of an isotropic suspension of rods, but that these
changes significantly affect the extent of clustering and therefore the
position of the percolation threshold.
\par
We model the CNTBs as hard spherocylinders, i.e., cylinders of 
length $L$ and diameter $D$ capped by hemispheres.  
To make as few assumptions as
possible about the effect of depletion, we choose to
model the surfactant micelles explicitly rather than by employing an effective
potential.  Explicit representation has the advantage of automatically
including many-body effects, due to multiple exclusion of the same volume,
which can never be perfectly accounted for by a pair potential.
In the experiments, the diameters of CNTBs and micelles are 
comparable \cite{wang:2004,vigolo:2005}, and we therefore model the 
micelles as hard spheres of diameter $D$.  The range of the depletion
interaction does depend on the depletant size, but for a given
sphere volume fraction, a spot check at aspect ratio $L/D=10$
revealed only a 2\% change in the percolation threshold using
sphere diameters of $0.5D$ to $1.3D$.
\par
Similar mixtures have been the subject of other theoretical and 
simulation studies \cite{roth:2002,li:2005,savenko:2006}.
In particular, the phase diagram for low volume fractions and roughly equal
rod and sphere diameters was studied within the second virial approximation
by Matsuyama and Kato \cite{matsuyama:2001}.  Most other studies, however,
have been concerned with different regimes, either where the rods are in
a liquid crystalline, rather than isotropic phase \cite{schoot:2002}, or
where the rods are regarded as the depletant rather than the
spheres \cite{chen:2002,bolhuis:2003,oversteegen:2004}.
We believe the present work to be the first time that depletion-induced
percolation has been simulated in rod--sphere mixtures.
\par
Monte Carlo simulations were performed at fixed volume and number of
particles using displacements and (for rods) rotations of individual
particles.  The hard potential reduces the Monte Carlo
acceptance criterion to the rejection of any trial
move that generates an overlap between rods and/or spheres.
For efficient overlap detection, the simulation box was divided
into cells with edge length $l \gtrsim D$ \cite{vink:2005}.
A cubic simulation box with periodic boundary conditions was used and,
except where stated, the box length was $5L$.
A typical simulation involved $5\times10^5$ trial
moves per particle (sweeps) for equilibration and a further $5\times10^5$ for
sampling.  Overall alignment of the rods was monitored using the nematic order
parameter $S$, which is the largest eigenvalue of the tensor
${\bf Q}=(2N_{\rm r})^{-1}\sum_i^{N_{\rm r}} (3\hat{\bf u}_i \hat{\bf u}_i - {\bf I})$.
Here, $\hat{\bf u}_i$ is a unit vector along the axis of rod $i$,
${\bf I}$ is the identity matrix, and the sum
is over all $N_{\rm r}$ rods in the system.  $S$ is unity in a fully aligned
configuration and zero in a perfectly isotropic fluid.
\par
To define clusters of rods, a connectivity criterion is required.  We are
not concerned with details of electrical conductivity at
CNTB contacts, and regard two rods to be connected if their surfaces
approach closer than $0.2D$, which is equivalent to the line segments at the
spherocylinders' axes approaching closer
than $A=1.2D$.  This is a physically reasonable, though necessarily
arbitrary cutoff. A cluster
was counted as percolating if its particles were connected to their
own periodic images. Percolation of the configuration was tested every
20 sweeps. 
\par
\begin{figure}
\includegraphics[width=80mm]{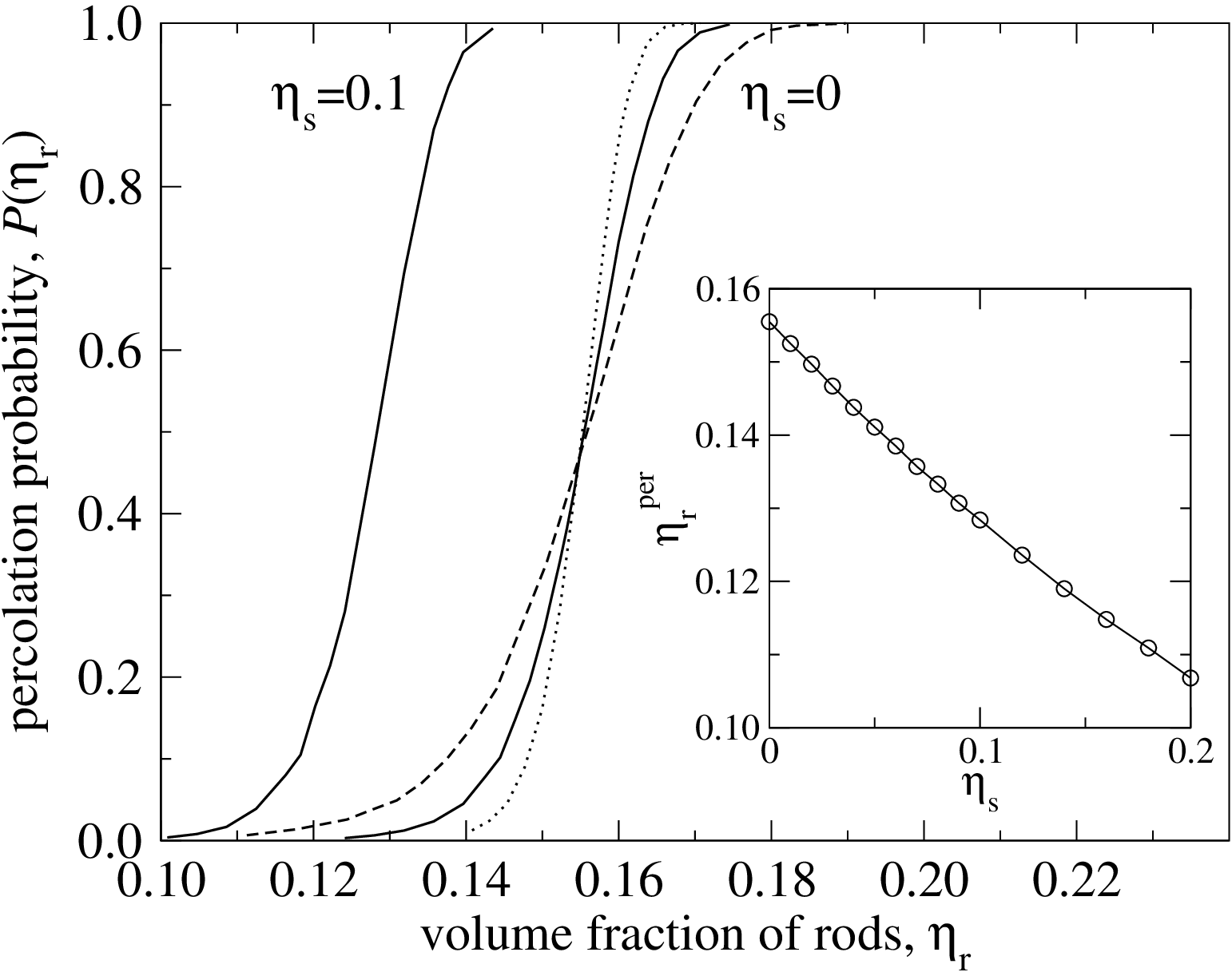}
\caption{\label{finitesize1} Solid lines: percolation probability as
a function of rod volume fraction for $L/D=6$ and two sphere volume
fractions $\etas$ in a box of edge $5L=30D$.
Dashed and dotted lines: equivalent results for $\etas=0$ only at
box lengths $20D$ and $45D$, respectively.  A typical uncertainty
in the probability is $0.004$.
Inset: percolation threshold as a function of sphere volume fraction
for $L/D=6$.}
\end{figure}

The aspect ratio of CNTBs can exceed 100, which is beyond the reach of
our simulations.  However, the emergence of long rod behaviour
in the rod--sphere mixture becomes clear by simulating increasing
$L/D$ starting from relatively small values.
Figure \ref{finitesize1} demonstrates the effect of depletion forces
on the percolation of rods with $L/D=6$.  The equilibrium probability
$P(\etar)$
of finding a percolating cluster is plotted as a function of the
rod volume fraction $\etar=N_{\rm r}\vh/V$, where $N_{\rm r}$ is
the number of rods, $\vh=\pi D^3(2+3L/D)/12$ is the volume of the
rod's hard core, and $V$ is the volume of the simulation cell.  Addition of
a volume fraction $\etas=0.1$ of spheres shifts $P(\etar)$
to lower rod volume fraction by $17.5\%$.

In an infinitely large system, $P(\etar)$ would rise instantaneously 
at the percolation threshold, but the simulations measure a sigmoidal 
curve whose width decreases with increasing cell size due to finite size effects,
as demonstrated for the pure rod case in Fig.~\ref{finitesize1}.
Nevertheless, the volume fraction at which $P(\etar)$ passes through 0.5 is 
almost independent of the cell size and we therefore adopt $P(\etat)=0.5$ as 
a robust definition of the percolation threshold $\etat$.
In general, above the sharp onset of percolation at $\etat$, the conductivity
of a network continues to increase with critical exponent $t$
\cite{stauffer:1994,foygel:2005} as $(\etar-\etat)^t$.
\par
As expected, $\etat$ is lower for longer rods,
as shown for a selection of $L/D$ ratios in Table \ref{data}.  Data
are given for rods with no depletant, and with
a volume fraction $\etas=0.03$ of spheres, which is comparable to the 
highest concentration of micelles used in \olcite{vigolo:2005}.
The fractional decrease $\delta$ in the percolation threshold due
to the spheres is $5.2\%$ at $L/D=4$,
rising gradually to $8.3\%$ by $L/D=25$ in contrast to
theoretical predictions of constant $\delta$ \cite{wang:2003}.
The increasing shift is likely to continue as the aspect ratio
rises towards that typical of CNTBs.  However, if a scaling regime
of the form $\etat\propto(L/D)^{-a}$ is entered for sufficiently
long rods, and the value of $a$ is not affected by the presence of spheres,
then the fractional drop in $\etat$ caused by a given volume fraction
of spheres would become independent of the aspect ratio.

\begin{table}

\caption{Percolation threshold for selected
aspect ratios $L/D$ at two sphere volume fractions $\etas$.
$\etat$ is the volume fraction of rods at percolation, and $\nt$ is the
average number of (rod) neighbors per rod.  $\delta$ is the fractional
decrease in $\etat$ caused by sphere volume fraction $\etas=0.03$.
\label{data}}
\begin{tabular}{clllllc}
\hline
& \multicolumn{2}{c}{$\etas=0$}
& & \multicolumn{2}{c}{$\etas=0.03$} & \\
\cline{2-3}
\cline{5-6}
$L/D$ & $\etat$ & $\nt$ & & $\etat$ & $\nt$ & $\delta$ \\
\hline
  4 & 0.178  & 1.61 & & 0.169  & 1.63 & 0.052\\
  6 & 0.156  & 1.54 & & 0.147  & 1.56 & 0.056\\
 10 & 0.124  & 1.45 & & 0.116  & 1.47 & 0.064\\
 17 & 0.0911 & 1.35 & & 0.0844 & 1.37 & 0.074\\
 25 & 0.0702 & 1.29 & & 0.0644 & 1.30 & 0.083\\
\hline
\end{tabular}
\end{table}

In fact, such a scaling has been predicted for ideal systems of 
freely interpenetrable rods.  In the hope of estimating percolation
thresholds without the need for detailed calculation, several quantities
have been suggested as approximate dimensional invariants at the
threshold.  Candidates included the number
of bonds per particle and the total excluded volume \cite{balberg:1984,pike:1974},
and it has been shown that the zero of the Euler characteristic correlates
well with percolation \cite{mecke:1991,mecke:2002}.  Although none of these approaches
works universally well for arbitrary particle shapes and mixtures, the
quantity $\rhot\langle\vex\rangle$ is an approximate invariant for ideal
rods \cite{foygel:2005}, where $\rhot$ is the number density at percolation
and $\langle\vex\rangle$ is the so-called excluded volume (orientationally
averaged) that would be inaccessible to the center of one particle due to the
presence of another if overlaps were forbidden.  Defining the total critical
volume by $\Vc=\rhot v$, where $v$ is the volume of an ideal rod, the invariant
becomes $\Vc\langle\vex\rangle/v$.  For long rods, $\langle\vex\rangle/v\propto L$
and hence $\Vc\propto1/L$.
\par
\begin{figure}
\includegraphics[width=75mm]{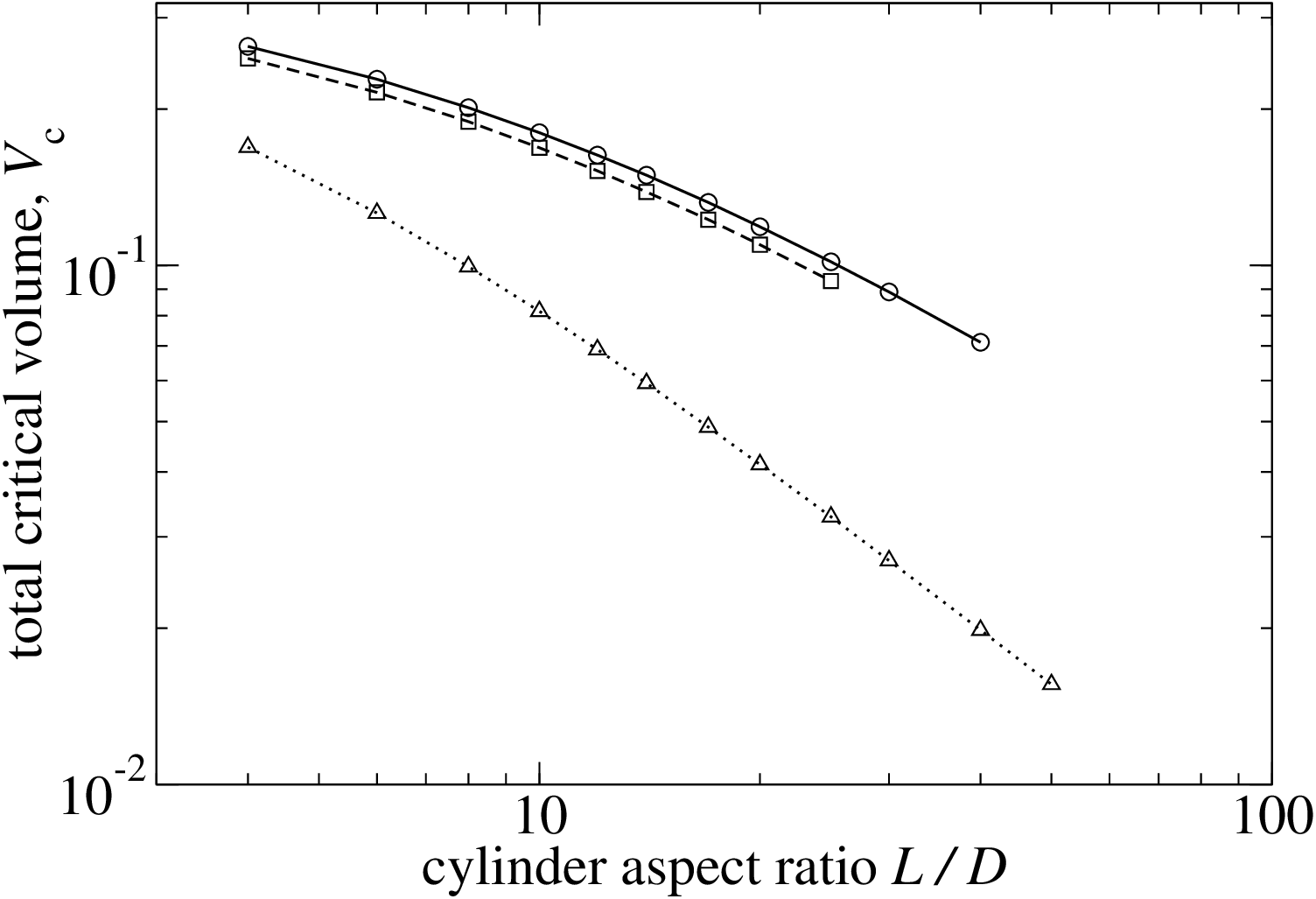}
\caption{Dependence of the total critical volume on the aspect ratio
of the rods for sphere volume fractions $\etas=0$ (solid line) and
$\etas=0.03$ (dashed line), and for the ideal system of freely interpenetrable
rods (dotted line).
\label{LoverD}}
\end{figure}

Ideal rods rapidly enter this scaling regime \cite{foygel:2005}, as shown
by the dotted line in Fig.~\ref{LoverD}.  Each evaluation of the percolation
probability was obtained from $10\,000$ uncorrelated and
randomly generated configurations
of line segments, using the same box size and connectivity criterion as
in the spherocylinder simulations.  The introduction of a hard
core induces local structure into the isotropic fluid and
rules out a large fraction of percolating ideal configurations,
raising the percolation threshold.
The total critical volume for hard rods with and without depletant spheres
is also shown in Fig.~\ref{LoverD}.
For the aspect ratios accessible to our simulations, no power-law scaling is
observed.  However, theoretical work \cite{leung:1991}
predicts that the effect of a hard core
is to postpone scaling to larger aspect ratios and that $1/L$ scaling is
eventually recovered.  Table \ref{data} shows a decline in the rate of increase
of $\delta$ with $L/D$, and in the scaling regime $\delta$ would level out to a
constant.  A linear extrapolation of $\delta$ to experimental aspect ratios
starting at $L/D=200$ \cite{vigolo:2005} therefore provides a crude upper bound to
the contribution of pure entropy in the CNTB--micelle experiments.  The
extrapolation predicts a lowering of 30--40\%, a sizeable effect but far smaller
than the experimental one.  The additional influence of direct energetic
interactions in the CNTB suspension could be tested by replacing the micelles by
a different depletion agent with the same excluded volume.

\begin{figure}
\includegraphics[width=70mm]{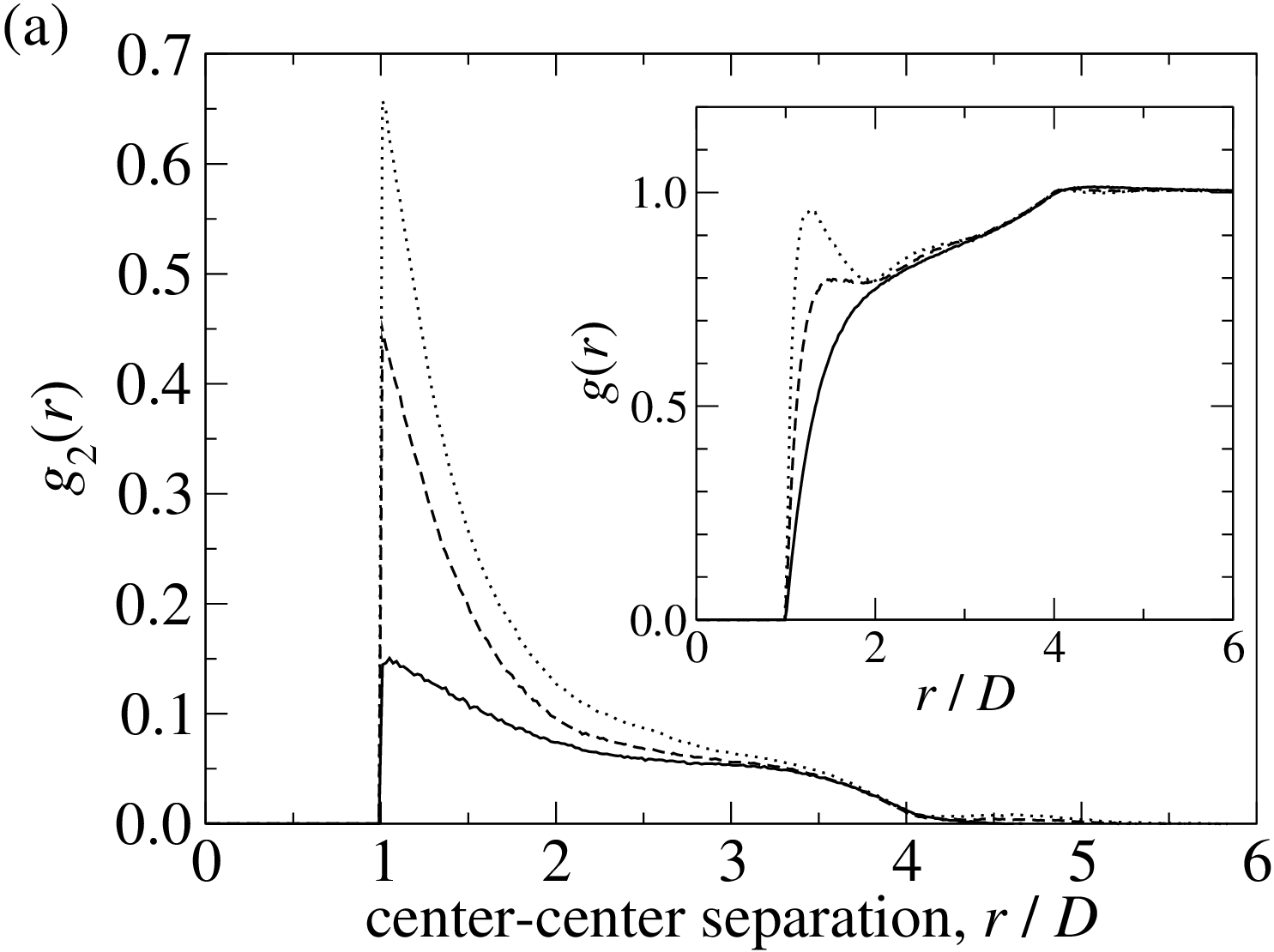}
\includegraphics[width=70mm]{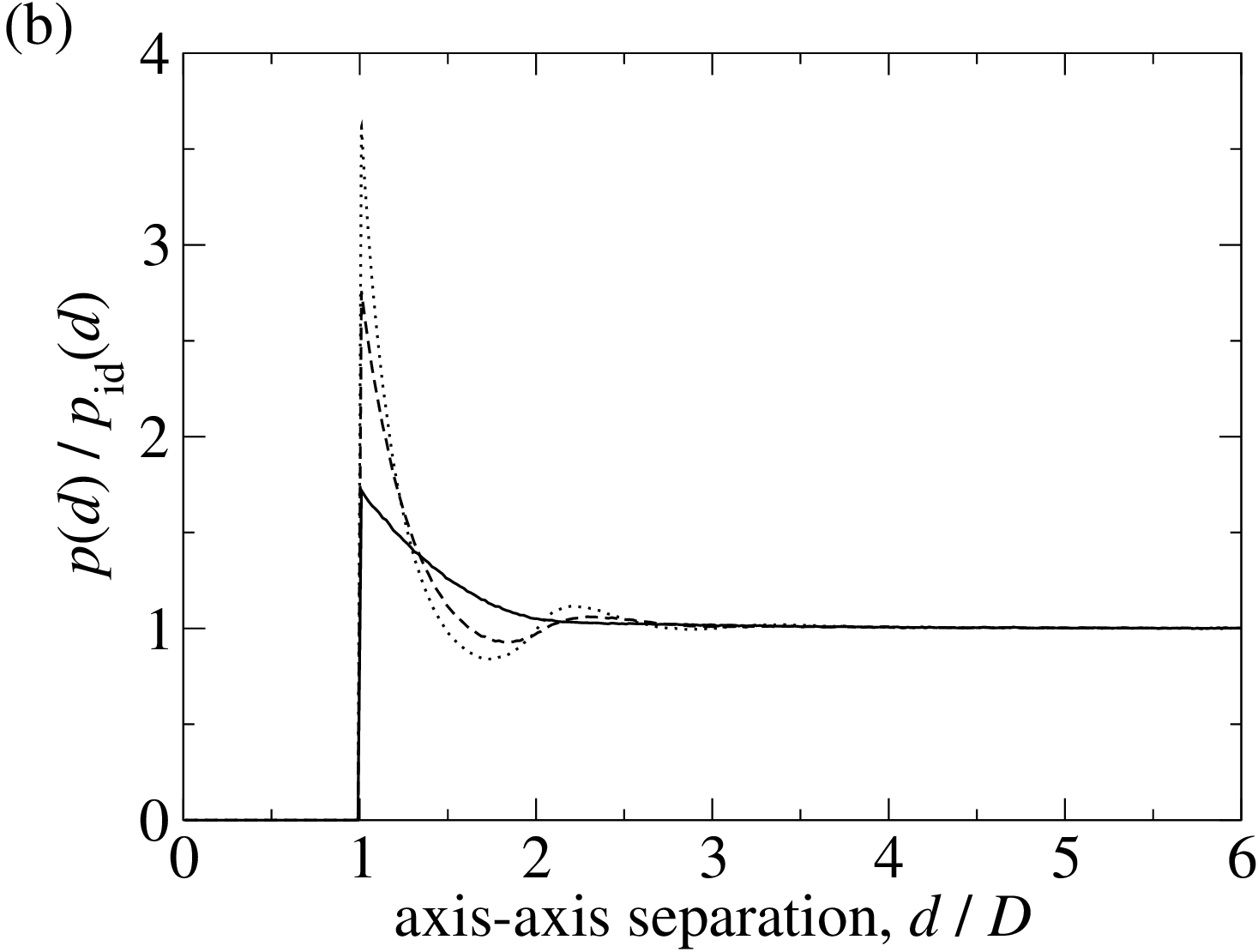}
\caption{
Local structure of a spherocylinder fluid with $L/D=6$
at volume fraction $\etar=0.142$.  (a) Orientational correlation function
$g_2(r)$, inset: radial distribution function, $g(r)$,
(b) distribution $p(d)$ of axis--axis separations normalized
by the ideal distribution $\pid(d)$.  Results for
three sphere volume fractions are shown: $\etas=0$ (solid lines), $\etas=0.1$
(dashed) and $\etas=0.15$ (dotted).
\label{structure}
}
\end{figure}

We now turn to structural characteristics of the fluid.  In all the simulations
the rods remain in the isotropic phase, which has no long-range order in either
the center-of-mass positions or the alignment of the rods.
However, the hard interactions give rise to local structure, which is affected
by the addition of hard spheres.  Local alignment is revealed by the 
orientational correlation function
$g_2(r)=\langle P_2(\hat{\bf u}_i\cdot\hat{\bf u}_j)\rangle_r$,
where $P_2(x)=(3x^2-1)/2$ is the second Legendre polynomial, and the averaging
is restricted to pairs of rods whose centers are separated by $r_{ij}=r$.
As shown in Fig.~\ref{structure}(a), $g_2(r)$ decays to
zero in an isotropic fluid, but becomes positive when rods approach closely
enough to restrict each other's rotation.

The addition of hard spheres to the rods encourages nearby rods to align, since
alignment decreases the total volume excluded to the spheres by an amount
proportional to the length of the adjacent segment, thereby increasing 
the entropy
of the spheres at the expense of some orientational entropy of the rods.
Figure \ref{structure}(a) shows that the spheres greatly enhance alignment of
rods with nearby centers and (inset) that the centers are pushed closer
together.  The range of the effect is on the order of the
sphere diameter $D$ and decays to the underlying correlation, whose range
depends on $L$.  For sphere concentrations $\etas=0.03$ around those of
the micelles in \olcite{vigolo:2005}, there is very little increase in
alignment beyond the range $D$, and no significant bundling of spherocylinders 
was observed.  At higher $\eta_r$, the spread of depletion-induced alignment
lowers the bulk isotropic--nematic transition density \cite{savenko:2006},
but following the percolation line $\etat(\eta_s)$ (inset of Fig.~\ref{finitesize1}),
we find that the nematic is preempted by rod--sphere demixing
(around $\etas=0.22$ for $L/D=6$).
\par
To understand cluster formation and percolation, a crucial structural property
is the probability distribution $p(d)$ of shortest distances
$d$ between pairs of rod axes, since it is the criterion $d<A$ that defines
contact between two rods.  To reveal the effect of the hard potential on
the structure, we divide $p(d)$ by the equivalent distribution $\pid(d)$
in a system of ideal rods.  The ratio
$p(d)/\pid(d)$ is plotted in Fig.~\ref{structure}(b).  As with the first peak
in the radial distribution function of a simple fluid, the distribution of
axis--axis separations rises above the random distribution for small $d$.
The addition of hard spheres again has a large but short-ranged effect on the
distribution, pushing the rods together at any point of approach along their
lengths.  For the high sphere volume fraction of $\etas=0.15$
a small second peak emerges at $d\approx 2.2D$, indicating a weak correlation between
two contacts with an intervening rod or sphere.  The enhancement of
$p(d)$ by the spheres for separations $d<A$ is a highly localized adjustment
to the structure of the pure rod fluid, but increases the connectivity
enough to cause the cluster size to diverge at lower rod concentration.
\par
The average number $\nt$ of contacts per rod at
percolation (Table \ref{data}), decreases with the aspect
ratio, confirming that, as for ideal objects, this quantity is not
an invariant \cite{pike:1974,mecke:2002,foygel:2005}.  The
number of contacts is a sensitive function of the rod density, so
it is interesting
to note that $\nt$ has similar values at $\etas=0$ and $0.03$; the spheres
lower the percolation threshold, but they increase the number of
contacts per rod at the lowered threshold roughly to its percolation
value their absence.
The low $\nt$ for long rods show how sparsely connected the network
is at the onset of percolation.  The strength of the network continues to
increase sharply above $\etat$.
\par
Using a simple model of hard spherocylinders and spheres, we have
isolated the effect of depletion on
the percolation threshold of nanorod networks.
The spheres induce pronounced but short-ranged changes in the structure of
the isotropic rod fluid, including enhanced local alignment.
At depletant concentrations relevant to experimental work on suspensions
of carbon nanotubes and micelles \cite{vigolo:2005}, the alignment
is too weak to cause global nematic order or significant aggregation of rods.
Although pairs of particles extend less far when mutually aligned, depletion
enhances contacts at all positions along the rods, leading to larger clusters
and a decrease in the percolation threshold.  It should therefore be possible
to enhance the percolation of nanorods using depletants other
than micelles, such as polymer coils or latex spheres.  The fractional
decrease caused by a given volume
fraction of depletant gradually increases with the aspect ratio of the rods.
This prediction could, in principle, be tested if shorter nanorods can
be synthesized and characterized.

\acknowledgments

We are grateful to the DFG (Tr6/D5 and Emmy Noether Program), the MWFZ
and Churchill College, Cambridge for support.  We
thank K.~Binder, P.~Poulin, E.~Frey and T.~Gruhn for helpful discussions.

\end{document}